\documentclass[aps,prl,notitlepage,nofootinbib,superscriptaddress,reprint]{revtex4-1}
\usepackage{graphicx}
\usepackage[tbtags]{amsmath}
\usepackage{amssymb,amsthm,mathtools}
\usepackage{braket}
\usepackage{epstopdf,cancel,ulem}
\usepackage{epsf,latexsym,bbm,euscript}
\usepackage[colorlinks,bookmarks=false]{hyperref}

\def\r{H_{rest}}

\usepackage{comment} % comment-out piece of text
\usepackage{color} %needed for coloring of \todo etc.
\usepackage{cancel}

\begin{document}

\author{Carolyn E. Wood}
\author{Magdalena Zych}
\affiliation{Australian Research Council Centre for Engineered Quantum Systems, School of Mathematics and Physics, The University of Queensland, St Lucia, QLD 4072, Australia}

%Title 15 words max.
\title{Composite particles with minimum uncertainty in spacetime}

%Abstract 150 words max. (159 words at 09/11/20)
\begin{abstract}
Composite particles---atoms, molecules, or microspheres---are unique tools for testing joint quantum and general relativistic effects, macroscopic limits of quantum mechanics, and searching for new physics. However, all studies of the free propagation of these particles find that they delocalise into separate internal energy components, destroying their spatial coherence. This renders them unsuitable for experimental applications, as well as theoretical studies where they are used as idealised test masses or clocks.
Here we solve this problem by introducing a new class of states with minimal uncertainty in space-time that fully overcome the delocalisation. The relevant physics comes from minimising the uncertainty between position and velocity, rather than position and momentum, while directly accounting for mass as an operator. Our results clarify the nature of composite particles, providing a currently missing theoretical tool with direct relevance for studies of joint foundations of quantum and relativistic phenomena, which removes a roadblock that could limit near-future quantum tests using composite particles.
\end{abstract}
\maketitle

%Introduction ~500 words max.
Progress in experimental quantum technologies has allowed us to push the boundaries of quantum mechanics with progressively more complex quantum systems and over increasingly large distances and time scales. Quantum interference has been observed with composite particles (molecules) comprising 2000 atoms~\cite{Fein2019}, and coherence of spatial superpositions has been verified over tens of centimetres~\cite{Asenbaum:2017PRL} and tens of seconds~\cite{Xu:2019Science20sec}. This progress brings us closer to testing new regimes and phenomena in fundamental physics which require control over many degrees of freedom ---namely, tests of joint quantum and general relativistic phenomena~\cite{Zych2011, bose2017spin, marletto2017entanglement, doi:10.1080/23746149.2017.1383184, loriani2019interference}, precision cosmology and gravity~\cite{Gratta:2016PRLDistToMicrosphere, Geraci:2016AIforDM, Schmoele:2016}, and the potential limits of quantum mechanics~\cite{ref:Bassi2013, pino2018chip}. All such experiments are highly susceptible to loss of spatial coherence: a problem which will only grow as the internal complexity and scale is increased.

The issue of spatial coherence loss will be particularly  detrimental to tests of relativistic gravity effects in quantum systems that aim to probe time-dilation effects on quantum coherence ~\cite{Zych2011, Zych:2012, bushev2016single, zych2018gravitational, Roura2020, Sonnleitner2018, Schwartz:2019hsn, SchwartzGiulini:2019,Pikovski2015, PikovskiTime2017, korbicz2017information}. Such experiments are referred to as clock-interference or quantum twin paradox tests---since small composite particles are, in relativity, a model of an ideal clock. Thus, quantum composite particles are considered idealized quantum clocks.

Currently, it is apparent that we are missing something in our understanding of the free propagation of composite particles. Theoretical studies of this scenario~\cite{Pang2016, Orlando2017, Anastopoulos2018} have found they delocalise into separate internal energy components, each travelling at different speeds. The same effect was discovered for dynamically-boosted particles~\cite{Paige2020}. This behaviour is contrary to even our most na\"{i}ve understanding of atoms and molecules as cohesive entities in the `real' world, where we expect them to have, at least in principle, well-localised spacetime trajectories. If this were unavoidable behaviour it would also upset the current theoretical paradigm, casting doubt on the suitability of composite particles as idealised clocks and test masses in quantum physics. It would be detrimental for both the above mentioned tests of fundamental physics, and for generic future experiments and metrology schemes with composite quantum systems.

In this work, we introduce a new class of quantum states and prove that they provide the optimal way to prepare composite particles to fully avoid the delocalisation problem, and the related loss of spatial coherence. These states are also the correct description of idealized quantum clocks following semi-classical trajectories. We show that the correct theoretical approach required to discuss limitations on the spacetime trajectories of composite quantum particles is to introduce a new uncertainty principle for position and velocity which includes mass as an operator. We then show that the quantum states which minimise the new inequality propagate coherently in spacetime, transform covariantly under boosts, and can be experimentally prepared in harmonic traps.

\subsubsection*{Phase vs configuration space of composite particles} 

Because of the mass-energy equivalence which entails that internal energy contributes to a particle's mass~\cite{Einstein:1905, Einstein:1907, PhysRevLett.4.341, Greenberger:1970a, Greenberger:1970b, Greenberger:1974}, phase space and configuration (position-and-velocity) space for composite particles are not related trivially. The internal mass-energy of a bound system---even at low centre of mass (CoM) energies---has a spectrum, and an inherent uncertainty. Failure to account for this fact can lead to inconsistent results, as shown in refs~\cite{sonnleitner2017will, Sonnleitner2018}. This also means that states of composite particles which propagate semi-classically in phase space will in general not have semi-classical propagation in spacetime. For example, Gaussian states---be they coherent or squeezed---as used in refs~\cite{Pang2016, Orlando2017, Anastopoulos2018} have minimum uncertainty in phase space, and thus do not have semi-classical spacetime trajectories.  

The states that will propagate semi-classically in spacetime must be defined from a configuration space (position and velocity) version of an uncertainty principle. Yet, despite extensive research on uncertainty principles of various types~\cite{ColesRev2017, Maccone2014, Franke2004}, motivated by their utility for minimising noise in precision experiments~\cite{Caves1981, Branciard2013}, uncertainty principles for configuration space variables have only been studied for structureless particles \cite{AlHashimi2009,Kowalski2018}. For composite particles, where mass is an operator, the problem has not been addressed. In order to find the required position-velocity uncertainty we first need the velocity operator for composite relativistic particles, which we introduce below.

\subsubsection*{Low-energy composite particles}
A composite particle can be described in a tensor product Hilbert space $\mathcal{H} = \mathcal{H}_{int} \otimes \mathcal{H}_{ext}$, where $\mathcal{H}_{int}$ is the Hilbert space describing the states of the internal degrees of freedom (DoFs) and $\mathcal{H}_{ext}$ those of the external ones (i.e.~the CoM states). The relativistic Hamiltonian (see also Methods) is
\begin{equation} \label{eq:relHam}
H=\sqrt{-g_{00}(c^2 p_j p^j +\r^2 c^4)},
\end{equation}

In the low-energy regime the Hamiltonian of a composite particle in the homogeneous gravitational field $g$ reads~\cite{ZychThesis, ZychBrukner2018, Anastopoulos2018, SchwartzGiulini:2019},
\begin{equation}\label{eq:Hamiltonian}
\hat{H} = \hat{M}c^2 + \frac{\hat{p}^2}{2\hat{M}} + \hat{M}gx,
\end{equation}
where $\hat{M} = m_0\hat{\mathbb{I}} + H_{int}/{c^2}$, with $m_0$ the ground state of the mass-energy (its `rest mass' parameter), $H_{int}$ describing the energy levels of the internal states, and $c$ the speed of light. Operators $\hat{x}$ and $\hat{p}$ are the position and momentum of the CoM degree of freedom. They satisfy the canonical commutation relation and act on $\mathcal{H}_{ext}$, while $\hat{M}$ acts on $\mathcal{H}_{int}$.

The form of the velocity operator, $\hat{v}= -\frac{i}{\hbar}\left[\hat{x},\hat{H}\right]$, will depend on the form of the Hamiltonian. The relativistic $\hat{v}$ takes the form
\begin{equation}\label{eq:relveloperator}
\hat{v} =\frac{\hat{p}c^2}{\sqrt{\hat{M}^2c^4+\hat{p}^2c^2}}
\end{equation}

At low energies, Eq.~\eqref{eq:Hamiltonian} is the relevant Hamiltonian, and Eq. \eqref{eq:relveloperator} reduces to
\begin{equation}\label{eq:velocityoperator}
\hat{v} \approx \frac{\hat{p}}{\hat{M}}
\end{equation}
The velocity operator is explicitly Hermitian, since all $\hat{x}, \hat{p}, \hat{H}$, and $\hat{M}$ are Hermitian. Eq. \eqref{eq:velocityoperator} stems from the canonical commutation relation for $\hat{x}$ and $\hat{p}$, and $\hat{x}$ and $\hat{p}$ each commute with $\hat{M}$ as they act on different Hilbert spaces. We also note that Eq.~\eqref{eq:velocityoperator} remains unchanged for any Hamiltonian that differs from Eq.~\eqref{eq:Hamiltonian} by a position dependent potential.

\subsubsection*{Position and velocity uncertainty and minimising states}
For any two arbitrary quantum observables, the minimum uncertainty states (MUSs) are those which minimise the Schr\"{o}dinger-Robertson uncertainty inequality---a stronger formulation of the more familiar Heisenberg-Robertson inequality carrying additional covariant terms~\citep{Schrodinger1930}. All such `generalised intelligent states'~\citep{Dodonov1980} are unitarily equivalent to the squeezed coherent states \cite{Stoler1970}.

For our scenario, we need to find the states which minimise the Schr\"{o}dinger-Robertson inequality for position and velocity:
\begin{equation}\label{eq:XVinequality}
(\Delta x)^2 (\Delta v)^2 - (\Delta xv)^2 \geq \frac{1}{4}\left|\left\langle\left[\hat{x},\hat{v}\right]\right\rangle\right|^2
\end{equation}
where the right hand side of Eq.~\eqref{eq:XVinequality} at low energies is $\left[\hat{x},\hat{v}\right] \approx \frac{i \hbar}{\hat{M}}$.%, and $m$ are real, positive eigenvalues.

The states that minimise Eq.~\eqref{eq:XVinequality} are solutions to the eigenvalue equation $\left(\mu\hat{a}_{\hat{M}} + \nu\hat{a}^\dagger_{\hat{M}} \right)\ket{\Psi} = z_{\hat{M}} \ket{\Psi}$, where $\mu,\nu,z \in \mathbb{C}$ and $|\mu|^2 - |\nu|^2 = 1$, $z_{\hat{M}} = \sqrt{\frac{\hat{M}}{2 \hbar}} z$, and $\hat{a}_{\hat{M}} = \sqrt{\frac{\hat{M}}{2\hbar}}\left(\hat{x} + i\frac{\hat{v}}{\Omega}\right)$. The general form of the solution is $\ket{\Psi} = \sum_m c_m \ket{\psi_m}\ket{m}$, where $\ket{m}$ is an eigenstate of $\hat{M}$ and $\ket{\psi_m}$ is the CoM state that explicitly depends on $m$. As a result, the full state $\ket{\Psi}$ exhibits entanglement between the internal and the centre-of-mass DoFs.

In the position representation, the minimising state $\Psi(x)$ for position and velocity uncertainty, which includes mass as an operator, has the form 
\begin{equation}\label{eq:minuncertwavefunc}
\sum_m c_m \psi_m(x)\ket{m} = \sum_m \frac{1}{\sqrt{\mathcal{N}_m}} e^{\frac{m}{2\hbar}\left[-\frac{\alpha}{\beta}(x- \frac{z}{\alpha})^2 + i\Im\left[\frac{z^2}{\alpha \beta}\right]\right]}\ket{m},
\end{equation}
where $\Im[\cdot]$ denotes the imaginary part of a complex number, $\alpha := (\mu+\nu)$, and $\beta := (\mu-\nu)$. Full derivation, including the normalization factor $\mathcal{N}_m$, can be found in the Methods section.

We compare these new states to a Gaussian state, such as would minimise the Schro\"{o}dinger-Robertson inequality in phase space:
\begin{equation}\label{eq:Gaussian}
\psi_G(x) = \frac{1}{\sqrt{\mathcal{N'}}} e^{\frac{1}{2\hbar}\left[-\frac{\alpha'}{\beta'}(x- \frac{z'}{\alpha'})^2 + i\Im\left[\frac{z'^2}{\alpha' \beta'}\right]\right]},
\end{equation}
with $\mathcal{N}'$ the normalisation factor. Importantly, the components of our MUS are mass-dependent and, in particular, have peak momentum  $m\Im\left[\frac{z}{\beta}\right]$ (for $\frac{\alpha}{\beta} \in \Re$) and thus mass-independent peak velocity. In contrast, $\psi_G(x)$ is independent of the mass---e.g. has peak momentum $\Im\left[\frac{z'}{\beta'}\right]$ (for $\frac{\alpha'}{\beta'} \in \Re$)---and thus mass-dependent propagation velocity, directly following from the fact that it minimises an uncertainty in phase space rather than configuration space. In the following sections, we explore the properties of our new class of states and compare them to the properties of the Gaussian states currently accepted as semi-classical states of free quantum particles. 

\subsubsection*{Particle propagation}
To obtain the propagated states, we use the path integral approach outlined in the Methods section. 

\begin{figure}[h!]
  \includegraphics[width=0.46\textwidth]{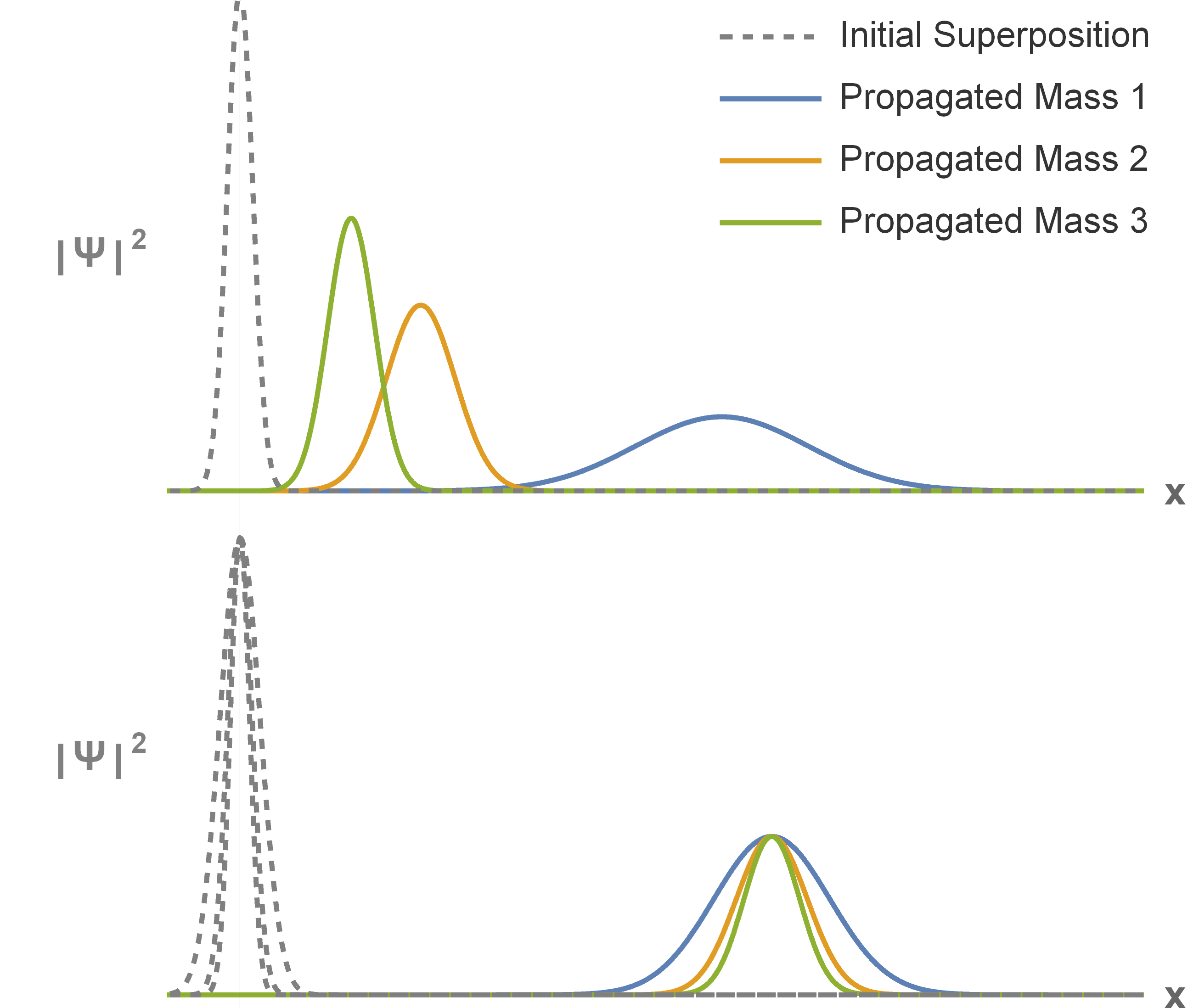}
  \caption{Propagation of a generic Gaussian state in an equal superposition ($\alpha_i\equiv1/\sqrt{3}$) of mass-energies (top) and the propagation of our position-and-velocity minimum uncertainty state (bottom). Initial states at $t=0$ (dashed grey lines), final state at $t=5$ time steps (solid coloured lines for each mass component, natural units). In the Gaussian state the mass components become separated and spread out at different rates. In our minimum uncertainty state the components propagate together for all times and spread at the same rate.\label{fig:superpositions}}
\end{figure}

We first analyse the propagation of a particle in a Gaussian state $\psi_G(x)$, Eq.~\eqref{eq:Gaussian}, with $\alpha,\beta, z \in \Re$, whose mass-energies are in a generic superposition:
\begin{equation}\label{eq:GenSup}
%\ket{\Psi_G} = 
\psi_G(x)\left(\sum_{i=1}^N \alpha_i{\ket{m_i}}\right),%\frac{\ket{m_i}}{\sqrt{N}}\right)
\end{equation}
where $\sum_i|\alpha_i|^2=1$. The analytical form of the wave function is given in the Methods section, and the initial and the propagated states are shown in Fig. \ref{fig:superpositions}, top-panel. The centres of the mass components shift in time as $x_i=pt/m_i$, as expected from Eq.~\eqref{eq:Gaussian}. Each travel with a different velocity {$p/m_i$} as they all have the same initial momentum $p$ but different mass-energy $m_i$. This is exactly the delocalisation effect found in prior studies~\cite{Pang2016,Orlando2017,Anastopoulos2018}.

Furthermore, the squared position variance of the Gaussian state Eq.~\eqref{eq:Gaussian} for each mass evolves as $\frac{\sigma^2}{2} \left(1+\frac{t^2\hbar^2}{m_i^2\sigma^4}\right)$. Thus for the case $p=0$ (stationary, expanding wave-packets) the position variance of the entire state becomes
\begin{equation}
\Delta x^2_G(t) = \sum_i |\alpha_i|^2\frac{\sigma^2}{2} \left(1+\frac{t^2\hbar^2}{m_i^2\sigma^4}\right).
\end{equation}
 
We now analyse the propagation of our MUS, Eq.~\eqref{eq:minuncertwavefunc}. The initial state takes the form
\begin{equation}\label{eq:IntelligentSuperpos}
%\ket{\Psi_{I}} = 
\sum_{i=1}^N \alpha_i \psi_{m_i}(x)\ket{m_i}.
%\frac{1}{\sqrt{N}} \psi_{m_i}(x)\ket{m_i},
\end{equation}

The analytical form of the state is again given in the Methods section. When propagated, its mass components remain all centred at the same position $x=vt$ (with $v=\Im[\frac{z}{\beta}]$, cf.~Eq.~\eqref{eq:minuncertwavefunc}), as shown in Fig.~\ref{fig:superpositions} bottom-panel. 
The position variance of each mass-energy component of our MUS evolves as $\sigma^2_{MUS}(m_i, t) = \sigma^2_{MUS}(m_i, 0)\left(1+e^{-4r}t^2\right)$, where $ \sigma^2_{MUS}(m_i, 0) \propto 1/m_i$ (cf.~Eq.~\eqref{eq:minuncertwavefunc}) and $\cosh[r]\equiv \mu$.

Thus the position variance of the entire state reads
\begin{equation}
\Delta x^2_{MUS}(t) = \Delta x^2_{MUS}(0)\left(1+e^{-4r}t^2\right),
\end{equation}
where $\Delta x^2_{MUS}(0) = \sum_i|\alpha_i|^2\sigma_{MUS}^2(m_i,0)/2$.

If we set $\Delta x^2_{MUS}(0) = \Delta x^2_G(0)$, so that the Gaussian and our MUS state both begin with the same width, we find
\begin{equation}
\Delta x^2_{MUS}(t) \leq \Delta x^2_G(t),
\end{equation}
with equality holding for the case $\alpha_i=\delta_{ij}$ (Kronecker delta) for some $j\in{1,...,N}$. This shows that our MUS is in general more localised than a Gaussian state, even when the mass-dependent delocalisation does not play a role (initial momentum and velocity both $=0$). 

The wavepackets discussed above are simply a special case of the MUS described in Eq.~\eqref{eq:minuncertwavefunc}, with real parameters. States with complex parameters, in analogy to ref.~\cite{Yuen1983}, can exhibit an additional `contractive' behaviour at short times--- see Appendix. Similarly to the case above, our MUS contracts as one cohesive state with all internal components reaching minimum width after the same propagation time, while the mass-energy components of the generic Gaussian each undergo the contraction at different times as well as delocalising as seen above.

Below we quantify the extent of the delocalisation between the propagating mass-energy components in the Gaussian state that is avoided by our MUS. Denoting the ground state mass-energy $m_{g}$, and its velocity $v_{g}=p_0/m_g$, and some higher mass-energy $m_e = m_g + \frac{\Delta E}{c^2}$, with velocity $v_e=p_0/m_e$, the difference in the velocities up to order $1/c^2$ is 
$v_g-v_e\approx v_{g}\frac{\Delta E}{m_{g}c^2}$. 

Using a Strontium atom as an example, due to its stable excited state with $\frac{\Delta E}{\hbar} = 10^{15}\, \text{Hz}$ and a lifetime of $\approx 100$~s~\cite{Bober2015}, we will have $m_{g} \approx 10^{-25}\,\text{kg}$, and the laboratory source will determine the initial CoM velocity of the atoms. If $v_{g}$ is the most probable velocity corresponding to $T = 800$~K~\cite{Poli2005}, we find $v_g-v_e \approx 10^{-9}\,\text{m/s}$. This means that in a Gaussian state, after around $10^{-3}$~s, the peak separation of the internal mass-energy states will become comparable with the atom's de Broglie wavelength, which is here around $10^{-12}$ m, thus suppressing longitudinal coherence~\cite{Schaff2015}.

Analogous estimations can be made for a molecule. The variance in the molecule's CoM velocity arising due to a thermal distribution of its internal mass-energies in a high temperature $T$ limit, and up to order $1/c^2$, is $\Delta v \approx\sqrt{3N-6} \,v_g\frac{k_BT}{mc^2}$ (where $N$ is the number of atoms and $k_B$ is the Boltzmann constant). Taking as an example data from ref.~\cite{Eibenberger2013}: $N=810$, $m=1.7\cdot 10^{-23}$~kg, $T=600$ K, de Broglie wavelength of the CoM of the molecule $\lambda_{dB}= 5\cdot 10^{-13}$~m and its size $10^4 \lambda_{dB}$, we find that the delocalization of the CoM $\Delta v\cdot t$ would be of the order $\lambda_{dB}$ after $t= 0.02$~s and would be as large as the size of the molecule after $t=3.3$ minutes, where we consider the size of the molecule to be the benchmark for complete loss of longitudinal coherence.

\begin{figure}
	\includegraphics[width=0.47\textwidth]{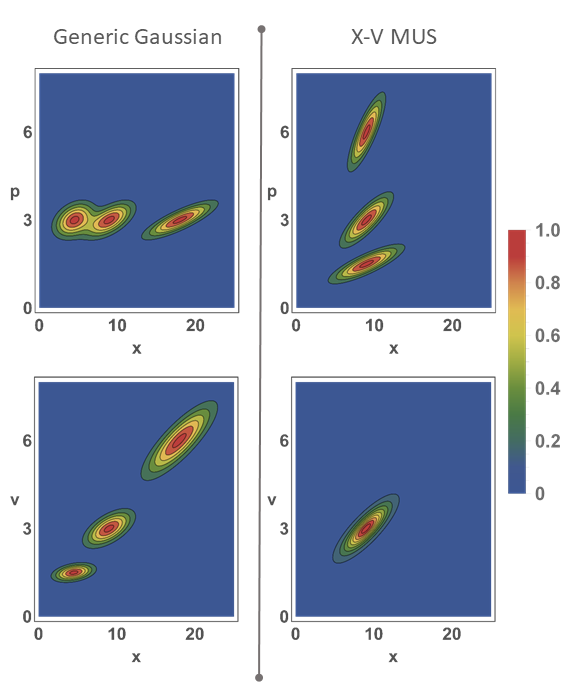}
	\caption{Wigner functions of time-evolved Gaussian state (left column) and our minimum uncertainty state (right column) in phase space (top row) and configuration space (bottom row). All states are initially (at $t=0$) centred at the origins of the plots and the plots show the state after $t=3$ time steps (in natural units). Each state is comprised of three masses $m=\{0.5,1,2\}$. The mass-energy components in the generic Gaussian state delocalise in both position and velocity due to different propagation speeds. In our minimum uncertainty state the mass-energy components remain localised in position and velocity, and the full state follows a semi-classical trajectory. In phase space, our minimum uncertainty state shows correlations between mass-energies and peak wavepacket momenta as expected from their common velocity. \label{fig:wignercontours}}
\end{figure}

\subsubsection*{Wigner functions in phase and configuration space}
Our new states do not only stay more localised in position, but also have a more defined space-time trajectory---i.e., path in configuration space---than the Gaussian states. We illustrate this using the Wigner quasi-probability distributions in phase space and in configuration space. 

The general form of the phase-space Wigner function for a mass-energy superposition state (see Methods section) is a sum of weighted Wigner functions for each mass-energy component, $W(x,p)=\sum_j \left|\alpha_j\right|^2 W^{(j)}(x,p)$\label{eq:wignersum}.

In the Methods section we derive a configuration space Wigner function, which takes the form: $\widetilde{W}(x,v)=\sum_j \left|\alpha_j\right|^2 W^{(j)}(x,m_j v)$. A similar function was used in ref.~\citep{Anastopoulos2018} in the context of the Weak Equivalence Principle for quantum particles.

Fig.~\ref{fig:wignercontours} shows results for time-evolved states from Eqs.~\eqref{eq:GenSup} and \eqref{eq:IntelligentSuperpos}. In configuration space, our MUS exhibits no separation of the mass states in either position or velocity, while the generic Gaussian state spreads out in both parameters. This demonstrates that our MUS indeed follows a semi-classical spacetime trajectory, in contrast to generic Gaussian states whose trajectory delocalises. In phase space, a generic Gaussian shows a spread in position, as observed in Fig.~\ref{fig:superpositions}, whereas our MUS remains localised in position and exhibits correlations between the individual mass-energy components and peak momenta.

\subsubsection*{Transformation of MUS under boosts}
From the perspective of composite particles as idealised clocks, a crucial characteristic of the MUSs introduced here is their covariant transformation under boosts.
This, combined with their cohesive propagation, means agents describing composite particles from different reference frames can apply relativistic transformations representing redshift (or, equivalently, time dilation) of the internal states of these particles, and will obtain the correct relation between their respective descriptions---in full agreement with classical intuition. 

To describe the required transformations, one needs an appropriate boost generator for composite particles. At low energies it is $e^{\frac{i}{\hbar}v(\hat{p}t - \hat{M}x)}$~\cite{ZychThesis,ZychGreenberger2019}; see also~\citep{Paige2020}.
Boosted from rest, the MUS yields an MUS moving with velocity $v$, as in Eq.~\eqref{eq:IntelligentSuperpos}. This is contrary to a boosted mass-superposition in a Gaussian state: A stationary Gaussian state from a moving reference frame has different peak momenta for the different mass-components, and thus differs from the state in  Eq.~\eqref{eq:GenSup}, see the Methods section for derivations. On the other hand, in a generic Gaussian state a superposition of masses does not even propagate cohesively, and thus is not a suitable representation of an ideal a clock either. The above shows that the MUS states fill a gap in our theoretical understanding of composite quantum systems. Specifically, particles in these new states can be seen as relativistic quantum clocks following trajectories as localised as quantum theory allows, and whose internal states `measure' proper time along these trajectories.

\subsubsection*{Double-slit interference}
Previous studies \cite{Orlando2017,Pang2016} looked at how particles in superpositions of internal mass-energy states interfere in double-slit-type experiments. The initial CoM states were taken to be Gaussian and it was found that the internal states interfere at different points of the screen when the particle is in free fall due to the difference in propagation velocities of the mass-energy components. This results in a mixture of interference fringes which suppresses interference. It has been argued \cite{Pang2016} that this effect is the true physical reason for the gravitational decoherence discussed in \cite{Pikovski2015}.

We show here that this is not the case, and demonstrate that our new class of states is the correct description of the double-slit realisation of gravitational decoherence\cite{Pikovski2015}. Importantly, the decoherence still occurs---only the interference pattern is suppressed---despite the fact that our states do not delocalise. We thereby separate two different effects on the coherence of quantum particles caused by the quantised mass-energy: time dilation decoherence \cite{Pikovski2015} and delocalisation-related decoherence \cite{Pang2016}.

To model the double-slit interference, the initial state is taken to be a superposition of states centred at two different locations (slits): $\ket{\Psi} = \sum_m c_m \left(\ket{\psi^L_m} +\ket{\psi^R_m}\right)\ket{m}$. Evolving the state in time, including the homogeneous gravitational field in the plane of the screen, yields the particle probability distribution at the screen; see Methods for details.

\begin{figure}[h!]
  \includegraphics[width=0.5\textwidth]{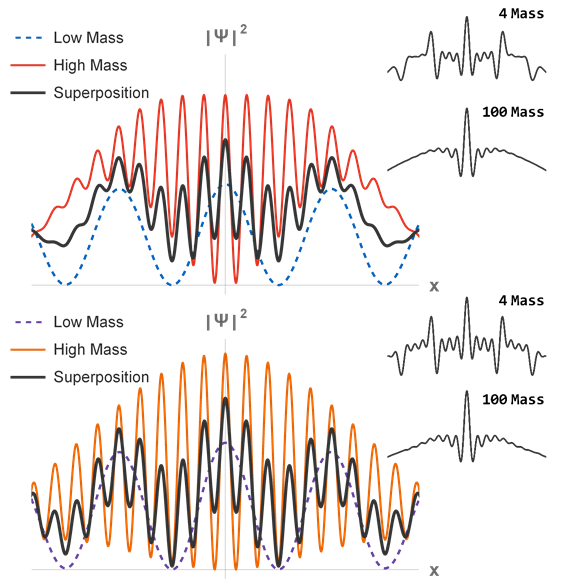}
    \caption{Double-slit interference of generic Gaussian superposition of two masses (top), and minimum uncertainty state (bottom). The mass-energy components of each (dotted and solid coloured lines) are explicitly shown along with the overall superpositions (thick, black line). Note the interference pattern for the MUS is more pronounced, while the Gaussian's interference washes out quickly as we move out from the centre. Insets: In the many-mass limit, both interference patterns approach a smooth classical distribution. The bright fringe in the centre is an artefact of our idealised case. We assume $N$ masses with a gap $\sim 1/N$, keeping mean mass and variance the same for all plots.
\label{fig:comparisoninterference}}
\end{figure}

Fig.~\ref{fig:comparisoninterference} plots the resulting interference for our MUS and a Gaussian initial state. 
Crucially, for the generic Gaussian, we ensure a common propagation velocity for the different mass-energy states, to take out the dominant effect of different arrival times, already studied in refs~\cite{Orlando2017,Pang2016} (this is done by considering the states spreading in the plane of the slits). The interference fringe modulations seen in Fig.~\ref{fig:comparisoninterference} are the two-mass limit of the gravitational decoherence. They do not vanish in either case, showing that different arrival times are not essential for this effect to appear. Moreover, the fringe modulations come from the time dilation, between different paths that interfere at the screen, that is encoded in the evolution of the internal mass-energy superposition. This path-dependent proper time difference hence affects the interference pattern as described in \cite{Zych2011,Pikovski2015}, and as expected from the complementarity between interference visibility and which-path information.

For larger, more complex systems, with internal states thermalised at high temperature T, recall that the velocity spread  for a generic Gaussian state is $v_g\sqrt{3N-6}k_BT/N\bar{m}$, where $v_g$ is the velocity associated with ground internal state, and $N\bar{m}$ is the total mass of the system with $\bar{m}$ the average mass of its constituents (taken to be atoms) and $N$ their total number. For large $N$ this becomes $\sim1/\sqrt{N}$ and thus in a macroscopic limit we recover the expected joint propagation of all the internal modes. In that limit, the beating in the interference pattern, Fig.~\ref{fig:comparisoninterference}, becomes more prominent,  fully washing away the interference~\cite{Pikovski2015}, see also refs~\cite{Zych2016, PikovskiTime2017}.

\subsubsection*{Discussion}
Our results show that the correct description of semi-classical states of composite quantum particles are position-and-velocity minimum uncertainty states. They fully avoid the delocalisation exhibited by Gaussian states and remain more localised as they spread---and consequently avoid major losses in spatial coherence. The new states provide the correct description of idealised quantum clocks, not only due to their lack of delocalisation but also due to their covariant transformation properties..

Furthermore, these new states can in principle be prepared straightforwardly in the laboratory: As the ground state of a harmonic potential for a massive particle is a Gaussian with squared width $\sigma^2\propto 1/m$, e.g.~\cite{Gersch1992}, a particle in a superposition of internal mass-energies, cooled down to the motional ground state of a harmonic trap that has a fixed frequency, would be prepared exactly in our MUS state Eq.~\eqref{eq:minuncertwavefunc}~\cite{Haustein2019}, with initial velocity given by the velocity of the trap in the laboratory reference frame. Traps with fixed frequency for the different internal states can be achieved for neutral particles where trapping is based on an induced dipole. For a generic wavelength of the trapping laser the effective harmonic potentials for the different internal states are generically different---due to different AC Stark shifts of the internal states. These can be made equal by choosing an appropriate (so-called `magic') laser wavelength  \citep{Derevianko2011,Ludlow2015}. In our context, one can thus choose the laser wavelength that provides fixed trap frequency for the different mass-energy states. Note that for traps of fixed stiffness the resulting states would neither be a Gaussian tensored with the internal states, nor one of our MUSs, see ref.~\citep{Haustein2019}.

We believe these new minimum uncertainty states can find applications in experiments testing interference of complex molecules~\cite{KialkaArndt:2019concepts, Fein2019}, nano- and microparticles~\cite{BatemanUlbricht:2014interferometry, MillenBarker:2015cooling, DelicKiesel:2019coherentscatter}, and in interference experiments with `quantum clocks'~\cite{Zych2011, bushev2016single, Roura2020} ---in which the delocalization effect, and associated loss of coherence, would become detrimental. Moreover, our results shed new light on fundamental differences between phase and configuration space for composite particles, which is particularly relevant to research on the equivalence principle in quantum mechanics~\cite{Viola1997, rosi2017quantum, Anastopoulos2018, ZychBrukner2018}. They will also find direct applications in theoretical studies of quantum models of ideal clocks at the interface with general relativity, such as \cite{Paige2020,rosi2017quantum,Sinha_2014,Castro:2017clocks,Khandelwal2020}. Finally, our study opens an avenue to further exploration of configuration space uncertainty principles, which may help address other fundamental  issues, such as limitations to high-precision timekeeping with quantum clocks due to couplings between internal and external DoFs~\cite{Sinha_2014, Castro:2017clocks, Paige2020, Khandelwal2020, SmithAhmadi:2019, Haustein2019}.

\section*{Methods}

\subsubsection*{Hamiltonian of a composite particle}
Recall first that the square of the relativistic four momentum $p^\mu$, $\mu=0,..,3$ is a relativistic invariant. It describes the energy of a particle in its rest frame~\cite{WeinbergGR1972}  $\r c^2=-\sum p^{\mu}g_{\mu\nu}p^{\nu}$, where $g_{\mu\nu}$ is a spacetime metric with signature $(-,+,+,+)$, and $c$ is the speed of light. In an arbitrary reference frame, the energy is $H\equiv cp_0$. Assuming a static symmetric metric we obtain Eq.~\eqref{eq:relHam}, where $p_jp^j\equiv\sum_{i,j=1,2,3}p^ig_{ij}p^j$. For a derivation of this dispersion relation from quantum field theory (as energy in a one-particle subspace) see~\cite{ZychThesis,CastroR2017,ZychBrukner2018,Anastopoulos2018}, for a derivation in a small-size limit of a bound system of N relativistic particles see~\cite{Zych2019}.

At low energies the relativistic Hamiltonian in Eq.~\eqref{eq:relHam} reduces to $\r+{p^2}/{2\r c^2}+\r\phi(x)/c^2$, with $\phi(x)$ denoting the gravitational potential. For a structureless particle: $\r\equiv mc^2$, where $m$ is the rest mass parameter. For a particle with internal DoFs, the rest energy comprises not only the masses of all the constituents but also the internal energies, as dictated by the relativistic mass-energy equivalence. For an atom or a molecule these include electronic and vibrational energies. We can thus write $\r=M_0c^2+H_{\mathrm{int}}$, where $M_0$ is the mass-energy of the system when the internal DoFs are in a ground state of rest energy; $M_0$ thus  defines the usual mass parameter familiar from the non-relativistic physics. The remaining $H_{\mathrm{int}}$ describes the dynamical part of the rest energy and can be identified as the internal Hamiltonian driving time evolution of the internal DoFs. For an atom, $H_{\mathrm{int}}$ can describe the electronic level structure, and for a molecule, the vibrational energy levels.
 
The low-energy limit $H\approx\r+{p^2}/{2\r c^2}+\r\phi(x)/c^2$ applies when the centre of mass energy is small enough to warrant the non-relativistic approximation but when the internal energy contributions to the kinetic and potential terms are non-negligible -- when mass-energy equivalence between internal energy and  mass of the system cannot be neglected. For this reason we denote the rest energy as $\r\equiv Mc^2$ and can write $H\approx Mc^2+{p^2}/{2M}+M\phi(x)$,
which is the Eq.~\eqref{eq:Hamiltonian} Hamiltonian in the main text. For the derivation up to $\mathcal{O}(1/c^2)$ in terms of an atom in a post-Newtonian metric see also~\cite{SchwartzGiulini:2019}.
\bigskip

\subsubsection*{Position-velocity minimum uncertainty states}
The minimum uncertainty states (MUSs) for two arbitrary quantum observables $\hat{X}$ and $\hat{Y}$ are the generalized intelligent states which minimise the Schr\"{o}dinger-Robertson uncertainty inequality~\citep{Dodonov1980},
\begin{equation}\label{eq:inequality}
(\Delta X)^2 (\Delta Y)^2 - (\Delta XY)^2 \geq \frac{1}{4}\left|\left\langle\left[\hat{X},\hat{Y}\right]\right\rangle\right|^2.
\end{equation}

Its MUSs are the solutions to the eigenvalue equation~\citep{Trifonov1994}
\begin{equation}\label{eq:genSSsolution}
(u \hat{A} + v \hat{A}^\dagger)\ket{\Psi} = z\ket{\Psi}, 
\end{equation}
where $z,u,v \in \mathbb{C}$ and $|u|^2 - |v|^2 = 1$,  $\hat{A} = \hat{X} + i\hat{Y}$, and  $\hat{A}^\dagger = \hat{X} - i\hat{Y}$.

As we are interested in MUSs for position and velocity, we define the operators in Eq.~\eqref{eq:genSSsolution} as
\begin{equation}
\hat{A} = \left(\hat{x} + i\frac{\hat{v}}{\Omega}\right)~; ~~ \hat{A}^\dagger = \left(\hat{x} - i\frac{\hat{v}}{\Omega}\right)
\end{equation}
where $\hat{X}$ in Eq.~\eqref{eq:inequality} becomes $\hat{x}$ and $\hat{Y}$ becomes $\hat{v}$, and with $\Omega$ an arbitrary parameter in units of frequency. We set $\Omega = 1$ for the remainder of this discussion.

The commutator on the right hand side of Eq.~\eqref{eq:inequality} is $\left[\hat{x},\hat{v}\right] = \frac{i \hbar}{\hat{M}}$, and
$\left[\hat{A}, \hat{A}^\dagger\right] = -2i\left[\hat{x},\hat{v}\right] = \frac{2\hbar}{\hat{M}}$.

It is then convenient to define operators $\hat{a}_{\hat{M}} := \sqrt{\frac{\hat{M}}{2 \hbar}}\hat{A}$ and $\hat{a}^\dagger_{\hat{M}} := \sqrt{\frac{\hat{M}}{2 \hbar}}\hat{A}^\dagger$ such that,
\begin{equation}\label{eq:commutator}
\left[\hat{a}_{\hat{M}},\hat{a}^\dagger_{\hat{M}}\right] = \frac{\hat{M}}{2\hbar} \left[\hat{A}, \hat{A}^\dagger \right] = \hat{\mathbb{I}}.
\end{equation}
This leads to a set of eigenvalue equations for the position and velocity case:
\begin{equation}\label{eq:ourintelleval}
\left(u\hat{a}_{\hat{M}} + v\hat{a}^\dagger_{\hat{M}} \right)\ket{\Psi} = z_{\hat{M}} \ket{\Psi},
\end{equation}
where $z_{\hat{M}} := z \sqrt{\frac{\hat{M}}{2 \hbar}}$.

As $\hat{M} = \sum_m m ~\hat{\Pi}_m$, where $\{\hat{\Pi}_m\}_m$ is a set of orthonormal projectors, we recast
\begin{equation}
\hat{a}_{\hat{M}} = \sum_m \left(\hat{a}_m \otimes \hat{\Pi}_m \right),
\end{equation}
where $\hat{a}_m = \hat{A} \sqrt{\frac{m}{2\hbar}}$ and, similarly, $z_m = z \sqrt{\frac{m}{2\hbar}}$.
Additionally, we can represent $\ket{\Psi} = \sum_m c_m \ket{\psi_m}\ket{m}$, so Eq.~\eqref{eq:ourintelleval} takes a more telling form:
\begin{equation}\label{eq:intelligent_m}
\sum_m \left(u\hat{a}_m + v\hat{a}^\dagger_m \right)c_m\ket{\psi_m}\!\ket{m}\!= \sum_m z_m c_m\ket{\psi_m}\!\ket{m},
\end{equation}
where the full MUS is made up of superposed states each with its own associated eigenvalue equation:
\begin{equation}\label{eq:singleeval}
\left(u\hat{a}_m + v\hat{a}^\dagger_m \right)\ket{\psi_m}\ket{m} = z_m \ket{\psi_m}\ket{m}.
\end{equation}
(Recall that $\ket{m}$ are eigenstates of the mass-energy of the particle.)

Since operators $\hat a_m$ satisfy the canonical commutation relations, Eq.~\eqref{eq:commutator}, each $\ket{\psi_m}$ is a squeezed Gaussian state with displacement parameter $\alpha_m=z_m$~\cite{Stoler1970,LoudonKnight1987}. 

In the position representation, the eigenstates of Eq.~\eqref{eq:intelligent_m} take the form of normalized wave functions:
\begin{equation}%\label{eq:minuncertwavefunc}
\psi_m(x) =  \frac{1}{\sqrt{\mathcal{N}_m}} e^{\frac{m}{2\hbar}\left[-\frac{\alpha}{\beta}(x- \frac{z}{\alpha})^2 + i\Im\left[\frac{z^2}{\alpha \beta}\right]\right]},
\end{equation}

where $\Im[\cdot]$ denotes the imaginary part of a complex number, $\alpha := (u+v)$, and $\beta := (u-v)$. The normalization factor is 
\begin{equation*}
\frac{1}{\sqrt{\mathcal{N}_m}} = \frac{\psi_m(0)}{\left|\psi_m(0)\right|}\left(\frac{m}{\pi\hbar} \Re[\frac{\alpha}{\beta}]\right)^{\frac{1}{4}} e^{\frac{m}{2\hbar}\left[-\frac{\Re[\frac{z}{\beta}]^2}{\Re[\frac{\alpha}{\beta}]} + \Re\left[\frac{z^2}{\alpha \beta}\right]\right]}.
\end{equation*}
\medskip

\subsubsection*{Path integral for composite particles}
The general form of a propagator is an integral over all possible trajectories for a given time interval~\cite{Feynman1948}. The propagator for our system is derived via the following expression:
\begin{equation*}
\begin{split}
&\braket{x_f, t_f, m'|x_i, t_i, m}\\ 
&\qquad \qquad = \bra{x_f,m'}e^{-\frac{i\hat{H}\Delta t}{\hbar}}\ket{x_i,m}\\
&\qquad \qquad = \bra{x_f, m'} e^{-\frac{i\Delta t}{\hbar}\left(\hat{M}c^2+\frac{\hat{p}^2}{2\hat{M}}+\hat{M}g\hat{x}\right)}\ket{x_i, m},
\end{split}
\end{equation*}
with Eq.~\eqref{eq:Hamiltonian} in the main text as the Hamiltonian, and $\Delta t = (t_f-t_i)$. The resulting expression is diagonal in the mass-energy components $\braket{x_f, t_f, m'|x_i, t_i, m} \equiv K_m(x_f, t_f; x_i, t_i)\delta_{m,m'}$, where $\ket{x_i, t_i, m} \equiv \ket{x_i, t_i}\ket{m}$.

Via the BCH (Zassenhaus) formula, and further noting that $\hat{M}$ commutes with both $\hat{x}$ and $\hat{p}$,

\begin{equation*}
\begin{split}
&\bra{x_f, m'} e^{-\frac{i\Delta t}{\hbar}\left(\hat{M}c^2+\frac{\hat{p}^2}{2\hat{M}}+\hat{M}g\hat{x}\right)}\ket{x_i, m} \\
&\qquad = \bra{x_f, m'} e^{-\frac{i\Delta t}{\hbar}mc^2} e^{-\frac{i\Delta t}{\hbar}\frac{\hat{p}^2}{2m}}e^{-\frac{i\Delta t}{\hbar}mg\hat{x}}\\
&\qquad \qquad \qquad \qquad e^{-\frac{i\Delta t^2 g \hat{p}}{2\hbar}} e^{\frac{i\Delta t^3 m g^2}{3\hbar}} \ket{x_i, m} \delta_{m,m'}\\
%&= \frac{1}{2\pi\hbar} \int dp ~e^{-\frac{i\Delta t}{\hbar}\left[\frac{p^2}{2m}-\frac{p(\Delta x)}{\Delta t}+mgx_i+\frac{\Delta t g p}{2}+\frac{\Delta t^2 m g^2}{6}+mc^2\right]} \delta_{m,m'}.
\end{split}
\end{equation*}
where, again, $\Delta t = (t_f-t_i)$, and $\Delta x = (x_f-x_i)$.

This yields the integral
\begin{equation*}
\begin{split}
= \frac{1}{2\pi\hbar} \int dp ~e^{-\frac{i\Delta t}{\hbar}\left[\frac{p^2}{2m}-\frac{p(\Delta x)}{\Delta t}+mgx_i+\frac{\Delta t g p}{2}+\frac{\Delta t^2 m g^2}{6}+mc^2\right]}\\
\qquad \times ~\delta_{m,m'}.
\end{split}
\end{equation*}
The solution to the integral gives $K_m$: our propagator for a particle with internal mass-energy $m$. The full propagator takes the form $\mathcal{K}(x_f, t_f; x_i, t_i)= \sum_mK_m(x_f, t_f; x_i, t_i) ~\hat{\Pi}_m$, where
\begin{widetext}
\begin{equation}\label{eq:masspropapp}
K_m(x_f, t_f; x_i, t_i) = \left(\frac{m}{2\pi\hbar i(\Delta t)}\right)^{\frac{1}{2}} e^{- \frac{imc^2(\Delta t)}{\hbar} \left[1-\frac{(\Delta x)^2}{2c^2 (\Delta t)^2}+\frac{g}{2c^2}(x_f+x_i)+\frac{g^2}{24c^2} (\Delta t)^2\right]}.
\end{equation}
\end{widetext}

The propagator is applied by convolving it with an initial wave function $\psi(x_i,t_i)$ to yield the final state (where we drop the subscript `f' for final state from here on for clarity):
\begin{equation}
\Psi(x,t) = \int dx_i\;\mathcal{K}(x,t;x_i,t_i)\Psi(x_i,t_i).
\end{equation}

The general form of the propagated MUS reads:
\begin{widetext}
\begin{equation}\label{eq:propIntell}
\psi_{MUS}(x,t) = \frac{1}{\sqrt[4]{\frac{\pi\hbar}{m\Omega}} \sqrt{1+ie^{-2r}t\Omega}} e^{\left[-\frac{m\Omega}{2\hbar}\frac{e^{-2r}(x-v_0t)^2}{1+e^{-4r}t^2\Omega^2}-\frac{r}{2} -\frac{imc^2t}{\hbar} \left(1+\frac{1}{2c^2t}\frac{-2v_0 x+v_0^2t-e^{-4r}x^2t\Omega^2}{1+e^{-4r}t^2\Omega^2}\right)\right]}
\end{equation}

whereas the general form of a propagated Gaussian state reads:

\begin{equation}\label{eq:propGauss}
\psi_G(x,t) = \frac{1}{\sqrt[4]{\pi}\sqrt{\sigma}\sqrt{1+\frac{it\hbar}{m\sigma^2}}}~e^{ \left[-\frac{\left(x-\frac{p}{m}t\right)^2}{2\sigma^2\left(1+\frac{t^2\hbar^2}{m^2\sigma^4}\right)}-\frac{imc^2t}{\hbar}\left(1+\frac{1}{2mc^2t}\frac{-2px+\frac{p^2t}{m}-\frac{x^2\hbar^2t}{m\sigma^4}}{1+\frac{t^2\hbar^2}{m^2 \sigma^4}}\right)\right]}
\end{equation}
\end{widetext}

\subsubsection*{Wigner representation}

Wigner quasi-probability distributions allow us to compare the minimum uncertainty states with the generic Gaussian states in both phase space and in configuration (position and velocity) space.

For a state $\ket\Psi$ of the composite particle, the Wigner function is defined as
\begin{equation}\label{eq:wignertrace}
W(x,p) = \int \frac{d\xi}{2\pi} e^{ip\xi} \text{Tr}_m\left\lbrace\braket{x + \frac{1}{2}\xi|\Psi}\braket{\Psi|x-\frac{1}{2}\xi}\right\rbrace.
\end{equation}

Expressing the state as $\ket\Psi =\sum_i\alpha_i\ket{\psi_i}\ket{m_i}$, the partial trace over the mass-energy gives
\begin{equation*}
\text{Tr}_m\{\ket{\Psi}\bra{\Psi} \} = \sum_j \left|\alpha_j\right|^2 \braket{x+\frac{1}{2}\xi|\psi_j}\braket{\psi_j|x-\frac{1}{2}\xi},
\end{equation*}
leaving an overall function comprised of a convex combination of Wigner functions for each mass-energy component,

\begin{align*}\label{eq:wignerint}
W(x,p) &= \sum_j \left|\alpha_j\right|^2 \int \frac{d\xi}{2\pi} e^{ip\xi} \psi_j(x+\frac{1}{2}\xi)\psi^*_j(x-\frac{1}{2}\xi) \\
&= \sum_j \left|\alpha_j\right|^2 W^{(j)}(x,p).
\end{align*}                                                                                                                                                                                                                                                                                                                                                                                                                                                                                                                                                                                                                                                                                                                                                                                                                                                                                                                                                                                                                                                                                                                                            

The Wigner representation of the propagated Gaussian function, where $\psi_j$ is given in Eq.~\eqref{eq:propGauss}, reads
\begin{equation}\label{eq:GenphaseWigner}
W_G(x,p) = \sum_j \left|\alpha_j\right|^2 \frac{1}{\pi \hbar} e^{-\frac{\left(\frac{p t}{m_j}-x\right)^2}{\sigma ^2}-\frac{\sigma ^2 (p-p_0)^2}{\hbar ^2}}.
\end{equation}

Similarly, the Wigner function for our propagated minimum uncertainty state Eq.~\eqref{eq:propIntell}, where for simplicity we put $r=0$ and $\Omega=1$, is
\begin{equation}\label{eq:MUSphaseWigner}
W_{{MUS}}(x,p) = \sum_j \left|\alpha_j\right|^2\frac{1}{\pi \hbar}~e^{-\frac{m_j}{\hbar}\left[\left(-\frac{pt}{m_j}+x\right)^2 + \left(\frac{p}{m_j}-v_0\right)^2\right]} \\
%&= \sum_j \left|\alpha_j\right|^2\frac{1}{\pi \hbar}~e^{-\frac{m_j}{\hbar}\left[\left(-v t+x\right)^2 + \left(v-v_{0}\right)^2\right]}.
\end{equation}

For a configuration (position and velocity) space Wigner function we change variables in Eq.~\eqref{eq:wignertrace} to $\xi' = m \xi$:
\begin{equation*}
\widetilde{W}(x,v) = \int \frac{d\xi'}{2\pi m} e^{iv\xi'} \text{Tr}_m\left\lbrace\braket{x + \frac{\xi'}{2m}|\Psi}\braket{\Psi|x-\frac{\xi'}{2m}}\right\rbrace.
\end{equation*}

For one mass, the equation above gives
\begin{equation*}
\int \frac{d\xi'}{2\pi m_j} e^{iv\xi'} \psi_j(x+\frac{\xi'}{2m_j})\psi^*_j(x-\frac{\xi'}{2m_j})\equiv W^{(j)}(x,m_jv),
\end{equation*}
which is simply the Wigner function where momentum is non-trivially dependent on the individual mass energies, such that $v=\frac{p_j}{m_j}$, as expected.

Consequently, the full Wigner function is again a sum of Wigner functions each corresponding to a different mass-energy state,
\begin{equation}
%&= \sum_j \left|\alpha_j\right|^2 \int \frac{d\xi}{2\pi} e^{i**m_jv**\xi} \psi_j(x+\frac{1}{2m_j**}\xi)\psi^*_j(x-\frac{1}{2m_j**}\xi)\\
\widetilde{W}(x,v) = \sum_j \left|\alpha_j\right|^2 W^{(j)}(x,m_jv).
\end{equation}

The Wigner function for our MUS in configuration space is thus
\begin{equation*}
\widetilde{W}_{{MUS}}(x,v)=\sum_j \left|\alpha_j\right|^2 \frac{1}{\pi  \hbar} ~e^{-\frac{m_j}{\hbar}\left[(x-t v)^2+(v-v_0)^2\right]},
\end{equation*}
where we note its similarity to the phase space Wigner function for our MUS, Equation \eqref{eq:MUSphaseWigner}.

The $x$-$v$ Wigner function of the generic Gaussian is, similarly,
\begin{equation}
\widetilde{W}_G(x,v) = \sum_j \left|\alpha_j\right|^2 \frac{1}{\pi \hbar} e^{-\frac{\left(v t-x\right)^2}{\sigma ^2}-\frac{m_j ^2\sigma ^2 (v-v_{0j})^2}{\hbar ^2}},
\end{equation}
where $v_{0j} := p_0/m_j$.
\medskip

\subsubsection*{Double slit interference}
In Fig.~\ref{fig:comparisoninterference}, the interference of the generic Gaussian and our MUS are compared.
The specific initial state used for the generic Gaussian is
\begin{equation}
\frac{1}{\sqrt{\mathcal{N}}}\left(e^{-\frac{\Omega_G}{2\hbar}(x-L)^2}+e^{-\frac{\Omega_G}{2\hbar}(x+L)^2}\right)\otimes\sum_m c_m \ket{m}
\end{equation}
where $L$ is the slit distance, and in the specific Fig.~\ref{fig:comparisoninterference} case of only two masses, $c_m = \frac{1}{\sqrt{2}}$.

For our MUS the initial state is
\begin{equation}
\sum_m c_m \frac{1}{\sqrt{\mathcal{N}_m}}\left(e^{-\frac{m\Omega_{\text{mus}}}{2\hbar}(x-L)^2}+e^{-\frac{m\Omega_{\text{mus}}}{2\hbar}(x+L)^2}\right)\ket{m}
\end{equation}

The initial widths of the two functions are set such that $\Omega_{\text{MUS}} =  \frac{\Omega_G}{2}(\frac{1}{m_1} + \frac{1}{m_2})$, making the position variance equal for the two states.

The propagator is applied to both states as outlined earlier, adding a gravitational acceleration term.

We note that the only effect of gravity on all the studied wave-packets is to shift the entire interference pattern by a classical free-fall distance $-gt^2/2$ where $g$ is gravitational acceleration and $t$ the propagation time. The plots can thus be equivalently interpreted as centred at $z_0=0$ in a gravity-free case and at $z_0=-gt^2/2$ in the case where the interfering particle is subject to a homogeneous gravitational field along the screen at which the interference is observed (perpendicular to the initial velocity of the wavepackets).
\medskip

\subsubsection*{State transformations under boosts}

To compare the behaviour of our MUS with that of the generic Gaussian state under a boost, we first discuss the appropriate boost generator for the mass-energy operator formalism.

Despite working in the low-energy regime, with the external motion of the particle being essentially classical, the internal relativistic dynamics preclude the simple use of the Galilean boost with a single mass parameter \cite{Greenberger2001, ZychGreenberger2019}. Since for each mass-energy eigenstate the formalism reduces to the non-relativistic one $ e^{\frac{i}{\hbar}v(\hat{p}t - {m}x)}$, one can construct the boost operator as $\sum_m e^{\frac{i}{\hbar}v(\hat{p}t - {m}x)}\ket{m}\bra{m}\equiv e^{\frac{i}{\hbar}v(\hat{p}t - \hat{M}x)}$. 

Below we show how this boost generator arises when considering an infinitesimal Lorentz transformation and taking the appropriate low energy limit. %Maybe we can write the Galilei boost and then say that considering it should hold  for each m, it is straightforward to extended, however, for completeness we derive it below from appropriate coordinate transformations
Beginning with two reference frames $S$ and $S'$, an infinitesimal Lorentz boost with velocity $v$ transforms the spacetime coordinates as $x' = x + vt$ and $t' = t + \frac{vx}{c^2}$. A wave function $\psi(x,t)$ in the $S$ reference frame reads  $\psi(x',t') = \psi(x+vt,t+\frac{vx}{c^2})$  in the $S'$ frame.

For $v$ infinitesimal we further have $\psi(x+vt, t+\frac{vx}{c^2}) = \psi(x) + v\left(t\nabla\psi(x)+ \frac{x}{c^2}\frac{\partial}{\partial t}\right)\psi(t)$, then,
\begin{equation}\label{eq:operatorboost}
\psi(x',t') =  e^{\frac{i}{\hbar}v(\hat{p}t - \hat{M}x)}\psi(x,t),
\end{equation}
which uses $-i\hbar\nabla=\hat p$ and  $i\hbar\frac{\partial}{\partial t}=\hat H$, and in the low energy limit $\frac{\hat{H}}{c^2} \rightarrow \hat{M}$.
We note that $\hat{p}t - \hat{M}x$ is the boost generator for the central extension of the Galilei group~\cite{ZychGreenberger2019} and can also be obtained from the In\"{o}n\"{u}-Wigner contraction of the Lorentz (or Poincar\'{e}) group~\cite{Inoenue:1952}. 

Eq.~\eqref{eq:operatorboost} recovers the anticipated boost generator and its action on mass-energy subspaces, i.e.~$e^{\frac{i}{\hbar}v(\hat{p}t - \hat{M}x)}\psi(x,t) = \sum_m e^{\frac{i}{\hbar}v(\hat{p}t - m x)}\psi_m(x,t)\ket{m}$.

To find the boosted states, $\bra{x',t'}\hat{U}_{boost}\ket{\psi}$, where $\ket{\psi} = \int dx \psi(x,t)\ket{x,t}$:
\begin{align*}
\psi(x',t') &= \bra{\widetilde{x}} e^{\frac{i}{\hbar}vt\hat{p}-\frac{i}{\hbar}mv\hat{x}} \int dx ~\psi(x,t)\ket{x}\\
&= \int dx \bra{\widetilde{x}}e^{\frac{i}{\hbar}vt\hat{p}}e^{-\frac{i}{\hbar}mv\hat{x}}e^{\frac{i}{2\hbar}v^2mt}\ket{x} \psi(x,t) \\
&= e^{-\frac{i}{\hbar}mv({x'}+vt)+\frac{i}{2\hbar}v^2mt} \psi({x'}+vt,t),
\end{align*}

This boost is then applied to the MUS and the generic Gaussian (Eqs~\eqref{eq:propIntell} and \eqref{eq:propGauss}), with the choice that the peak velocity for the MUS and the peak momentum for the Gaussian state, respectively, are set to be zero.

Hence, for one mass, the boosted position-velocity MUS takes the form:
\begin{widetext}
\begin{equation}
\psi_{MUS}(x,t) = \frac{1}{\sqrt[4]{\frac{\pi\hbar}{m\Omega}} \sqrt{1+i e^{-2r}t\Omega}} e^{\left[-\frac{m\Omega}{2\hbar} \frac{e^{-2r}(x+vt)^2}{\left(1+ e^{-4r}t^2\Omega^2\right)}-\frac{r}{2}-\frac{imc^2t}{\hbar} \left(1+\frac{1}{2c^2t}\frac{2vx+v^2t-e^{-4r}x^2\Omega^2t}{1+e^{-4r} t^2\Omega^2}\right)\right]}
\end{equation}

Note here that when the factor $e^{-r}$ tends to zero, the imaginary part of the exponent will become $\frac{imc^2t}{\hbar} \left(1+\frac{vx}{c^2t}+\frac{v^2}{2c^2}\right)$, which is a Taylor series expansion of the Lorentz factor, to 2nd order. This shows that the internal DoFs of particles in our MUS undergo time dilation in accordance with classical relativity. Combined with their semi-classical trajectories, this corroborates our statement that these new states are the correct description of ideal quantum clocks.

The boosted generic Gaussian state is:
\begin{equation}\label{eq:mGaussian}
\psi_{mG}(x,t) = \frac{1}{\sqrt[4]{\pi } \sqrt{\sigma } \sqrt{1+\frac{i t \hbar }{m \sigma ^2}}} e^{\left[-\frac{(x+vt)^2}{2 \sigma ^2 \left(1+\frac{t^2 \hbar ^2}{m^2 \sigma ^4}\right)}-\frac{imc^2t}{\hbar} \left(1+\frac{1}{2mc^2t}\frac{2mvx+mv^2t-\frac{x^2\hbar^2t}{m\sigma^4}}{1+\frac{t^2 \hbar ^2}{m^2 \sigma ^4}}\right)\right]} 
\end{equation}

\end{widetext}

Comparing these two states with \eqref{eq:propGauss} and \eqref{eq:propIntell}, we can see that these individual mass-energy components have the exact same form. However, for a full mass-superposition state, we find that the MUS is covariant under the boost -- we get exactly a superposition corresponding to that obtained from \eqref{eq:propIntell}, e.g.~Eq.~\eqref{eq:IntelligentSuperpos} in the main text. 

On the other hand, each mass component of the generic Gaussian superposition obtained from Eq.~\eqref{eq:mGaussian} will have a different momentum $p_j=m_jv$:
\begin{widetext}
\begin{equation*}
\sum_j\alpha_j\psi_{m_jG}(x,t)\ket{m_j} = \sum_j\alpha_j\frac{1}{\sqrt[4]{\pi } \sqrt{\sigma } \sqrt{1+\frac{i t \hbar }{m_j \sigma ^2}}} e^{\left[-\frac{(x+\frac{p_j}{m_j}t)^2}{2 \sigma ^2 \left(1+\frac{t^2 \hbar ^2}{m_j^2 \sigma ^4}\right)}-\frac{i m_jc^2t}{\hbar} \left(1+\frac{1}{2m_jc^2t}\frac{2p_jx+\frac{p_j^2t}{m_j}-\frac{x^2t\hbar^2}{m_j\sigma^4}}{1+\frac{t^2 \hbar ^2}{m_j^2 \sigma ^4}}\right)\right]}\ket{m_j} 
\end{equation*}
\end{widetext}
which differs from a Gaussian state with a fixed peak momentum tensored with the internal mass-superposition state, $\psi_G(x,t)\alpha_j\ket{m_j}$, as in Eq.~\eqref{eq:GenSup} in the main text.
\bigskip

\textit{Acknowledgements.--}
We thank F.~Costa, A.~Kempf, T.C.~Ralph  and N.~Stritzelberger for discussions, and A.G.~White for helpful advice. This research was supported by ARC grants CE170100009, DE180101443, and the University of Queensland grant UQECR1946529; 
The authors acknowledge the traditional owners of the land on which UQ is situated, the Turrbal and Jagera people.

\appendix
\begin{widetext}
\section{Contractive States}
In the main text, we presented our position-velocity minimum uncertainty state (Equation (6) in the main text):
\begin{equation}\label{eq:MUSCont}
\Psi_{\text{MUS}}(x) = \sum_m \frac{c_m}{\sqrt{\mathcal{N}_m}} e^{\frac{m}{2\hbar}\left[-\frac{\mu + \nu}{\mu - \nu}(x- \frac{z}{\mu + \nu})^2 + i\Im\left[\frac{z^2}{(\mu - \nu)^2}\right]\right]}\ket{m}
\end{equation}
and the corresponding generic Gaussian (Equation (7)):
\begin{equation}\label{eq:GaussianCont}
\Psi_G(x) = \frac{1}{\sqrt{\mathcal{N'}}} e^{\frac{1}{2\hbar}\left[-\frac{\mu' + \nu'}{\mu' - \nu'}(x- \frac{z'}{\mu' + \nu'})^2 + i\Im\left[\frac{z'^2}{(\mu' - \nu')^2}\right]\right]} \otimes \sum_m c_m \ket{m},
\end{equation}
and used real parameters in each to perform the subsequent investigations of the behaviour of the new states.

However, if one includes complex parameters, one can find so-called contractive states---first identified by Yuen \cite{Yuen1983} for a single free mass. These states experience an initial decrease in their position variance upon propagation before expanding.  

Contractive states have been studied as a means of beating the standard quantum limit, more recently considered in a double-slit scenario (for a fixed mass)\cite{Viola2003}.
\medskip

For our analysis here, the relevant parameter to consider is $\xi :=\Im[\mu^*\nu]$. The contractive behaviour occurs when the quantity $\xi > 0$, while for any $\xi \leq 0$, the position variance increases monotonically. We take one of our states as in Eq. \eqref{eq:MUSCont}, and the generic Gaussian as in Eq. \eqref{eq:GaussianCont} both in superpositions of three masses, and examine their behaviour when this contractive property is present, see Figure \ref{fig:contractivestates}.

\begin{figure}[h!]
  \centering
  \includegraphics[width=\textwidth]{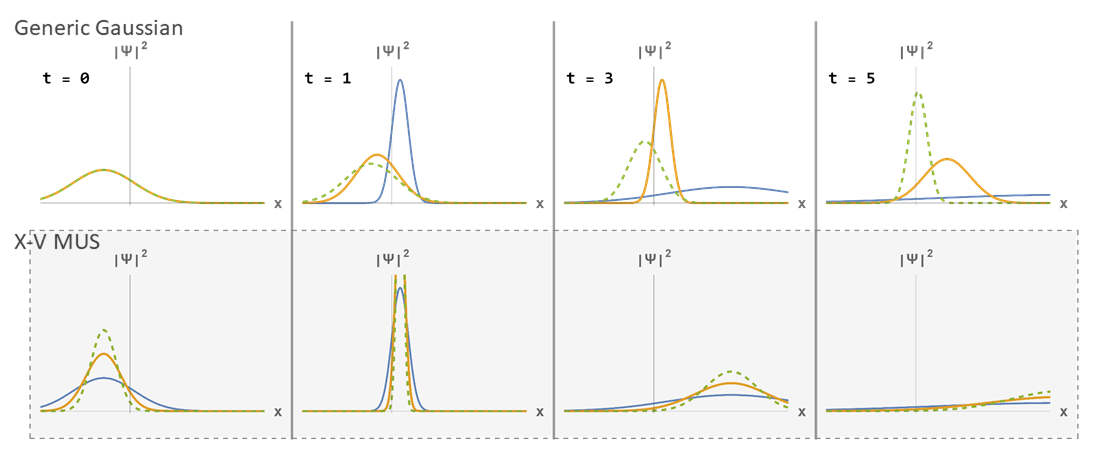}
    \caption{Position-space probability amplitudes for propagating states over four time slices. Both generic Gaussian (top row) and our MUS (bottom row, shaded), are in a superposition of three masses, the smallest mass being the blue, thin line, the largest the green, dashed line. Axes same scale for all plots. While in the Gaussian states different mass components exhibit the position-focusing at different times, in the MUS all internal states reach the the minimum position variance simultaneously. \label{fig:contractivestates}}
\end{figure}

As the mass components of the generic Gaussian superposition each travel at different velocities, they reach the point of contraction at different times. On the other hand, our MUS, with all components travelling at the same velocity, contracts as one cohesive entity. Figure \ref{fig:t1contractivestates} presents the MUS and the generic Gaussian state at two different times, chosen such that a specific mass component (here picked to be the middle of the three masses) is at its maximal contraction. While this time is the same for the MUS state, it increases with the mass for a Gaussian initial state.

\begin{figure}[h!]
  \centering
  \includegraphics[width=0.7\textwidth]{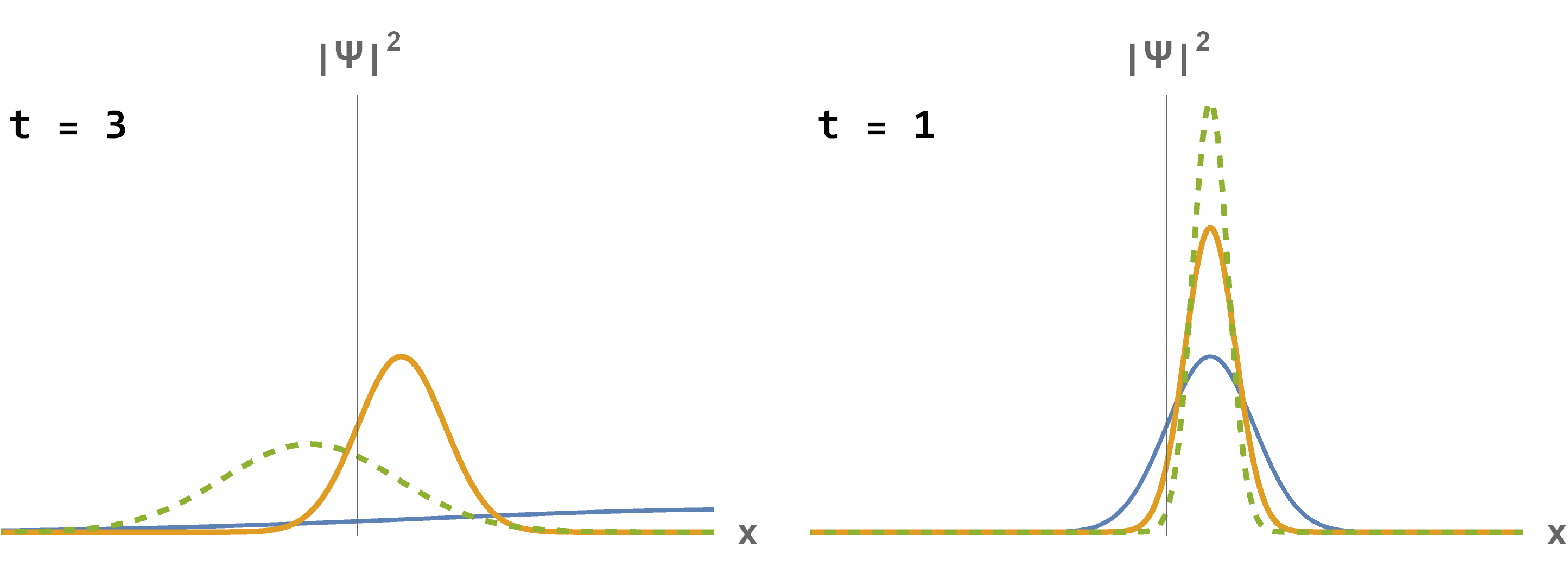}
    \caption{Position-space probability amplitude of the generic Gaussian (left) and our MUS (right) for different time slices, chosen such that maximal contraction occurs for a specific mass component (orange, thick line). The smallest mass is represented by the blue, solid line, the largest mass by the green, dashed line. Axes are adjusted from Figure 1, but with the same scale for both plots. Apart from the different times of maximal contraction for each mass component in the Gaussian state, which is absent in our MUS, the minimal widths for the same mass also differ between these two cases. This difference arises from the difference in the initial state of the mass components in the Gaussian state and in the MUS.    \label{fig:t1contractivestates}}
\end{figure}

In this way, one sees that while the contractive behaviour may be considered desirable for beating the standard quantum limit, only our states allow coherent display of this property for composite particles.
\end{widetext}

%\bibliographystyle{linksen}
%\bibliography{paper_literature}
%
%\end{document}

\providecommand{\href}[2]{#2}\begingroup\raggedright\endgroup

\end{document}